\documentclass[
a4paper,11pt,
amsmath,
amssymb,
accepted=2022-04-18]{quantumarticle}
\pdfoutput=1

\usepackage[utf8]{inputenc}
\usepackage[english]{babel}
\usepackage[T1]{fontenc}
\usepackage{amsmath}
\usepackage{lipsum}

\usepackage[numbers]{natbib}
\usepackage{graphicx}
\usepackage{dcolumn}

\usepackage{soul}
\sethlcolor{pink}

\usepackage[dvipsnames]{xcolor}
\usepackage{tikz}
\usetikzlibrary{arrows}
\usetikzlibrary{calc}
\usepackage[caption=false]{subfig}
\graphicspath{ {./images/} }
\usepackage{physics}
\usetikzlibrary{shapes.geometric}
\usetikzlibrary{positioning}

\usepackage{url}
\usepackage{hyperref}
\hypersetup{
    colorlinks=true,
    linkcolor=blue,
    citecolor=blue, 
    filecolor=magenta,      
    urlcolor=blue,
}

\usepackage{mleftright} 

\newcommand{\figref}[1]{\mbox{Fig.~\ref{#1}}}

\newcommand{\secref}[1]{\mbox{Sec.~\ref{#1}}}

\renewcommand{\eqref}[1]{\mbox{Eq.~(\ref{#1})}}

\newcommand{\figpanel}[2]{Fig.~\hyperref[#1]{\ref*{#1}(#2)}}
\newcommand{\figpanels}[3]{Fig.~\hyperref[#1]{\ref*{#1}(#2)-(#3)}}
\newcommand{\figpanelNoPrefix}[2]{\hyperref[#1]{\ref*{#1}(#2)}}

\usepackage{units}

\definecolor{Xorange}{HTML}{F5793A}
\definecolor{Ypurple}{HTML}{A95AA1}
\definecolor{Zblue}{HTML}{85C0F9}
\definecolor{plaq}{HTML}{0F2080}

\newcommand{\phn}[1]{XYZ$^2$ #1}

\begin{document}

\title{The XYZ$^2$ hexagonal stabilizer code}

\author{Basudha Srivastava}
\email[]{basudha.srivastava@physics.gu.se}
\affiliation{Department of Physics, University of Gothenburg, SE-41296 Gothenburg, Sweden}
\orcid{0000-0002-4972-4216}

\author{Anton Frisk Kockum}
\affiliation{Department of Microtechnology and Nanoscience, Chalmers University of Technology, SE-41296 Gothenburg, Sweden}
\orcid{0000-0002-2534-3021}

\author{Mats Granath}
\email[]{mats.granath@physics.gu.se}
\affiliation{Department of Physics, University of Gothenburg, SE-41296 Gothenburg, Sweden}
\orcid{0000-0003-3185-2014}

\date{}

\captionsetup[subfigure]{labelformat=empty}

\begin{abstract}

We consider a topological stabilizer code on a honeycomb grid, the ``XYZ$^2$'' code. The code is inspired by the Kitaev honeycomb model and is a simple realization of a ``matching code'' discussed by Wootton~\cite{Wootton2015}, with a specific implementation of the boundary. It utilizes weight-six ($XYZXYZ$) plaquette stabilizers and weight-two ($XX$) link stabilizers on a planar hexagonal grid composed of $2d^2$ qubits for code distance $d$, with weight-three stabilizers at the boundary, stabilizing one logical qubit. We study the properties of the code using maximum-likelihood decoding, assuming perfect stabilizer measurements. For pure $X$, $Y$, or $Z$ noise, we can solve for the logical failure rate analytically, giving a threshold of \unit[50]{\%}. In contrast to the rotated surface code and the XZZX code, which have code distance $d^2$ only for pure $Y$ noise, here the code distance is $2d^2$ for both pure $Z$ and pure $Y$ noise. Thresholds for noise with finite $Z$ bias are similar to the XZZX code, but with markedly lower sub-threshold logical failure rates. The code possesses distinctive syndrome properties with unidirectional pairs of plaquette defects along the three directions of the triangular lattice for isolated errors, which may be useful for efficient matching-based or other approximate decoding.

\end{abstract}

\maketitle


\section{Introduction}

In quantum computing, quantum bits, or qubits, are the basic unit of information. Whereas a classical computer acts on a single bit string, i.e., a list of 0's and 1's, a quantum computer can address states that are superpositions of bit strings, or even non-classical entangled bit strings. While this enhanced representability gives quantum algorithms the potential for significant speed-up compared to classical algorithms~\cite{Nielsen2010QuantumInformation, Georgescu2014QuantumSimulation, Montanaro2016QuantumOverview, Wendin2017QuantumReview, Preskill2018QuantumBeyond, McArdle2020QuantumChemistry, Bauer2020QuantumScience, Orus2019QuantumProspects, Cerezo2021VariationalAlgorithms}, it also makes quantum computing much more difficult to implement in practice. Whereas unavoidable errors due to noise in a classical computer are discrete bit flips, errors in a quantum computer are continuous, corresponding to arbitrary rotations on a unit (Bloch) sphere.

One way to cope with errors is to implement error correction~\cite{Shor1995SchemeMemory, Steane1996ErrorTheory, Gottesman1997StabilizerCorrection, Terhal2015QuantumMemories, girvin2021introduction}. In quantum computing, this can be done using stabilizer codes. These codes are quantum algorithms that turn a collection of noisy qubits into a single logical qubit which is less error-prone~\cite{Nielsen2010QuantumInformation}. The stabilizers of the code correspond to a set of commuting measurements on groups of qubits that project continuous errors to discrete errors on the code, which can be represented by Pauli $X$, $Y$, or $Z$ operations on individual qubits. Topological codes, such as Kitaev's surface code~\cite{Bravyi1998QuantumBoundary, Dennis2002TopologicalMemory, Kitaev2003Fault-tolerantAnyons, Raussendorf2007Fault-TolerantDimensions, Fowler2012SurfaceComputation}, are particularly attractive for practical purposes since they use stabilizers consisting of geometrically local qubit operations on a two-dimensional grid. At the same time, logical errors are exponentially suppressed with increasing linear dimension of the code, provided the error rate is below the error threshold of the code~\cite{ShorFaulttolerantComputation, Knill1998ResilientComputation, Dennis2002TopologicalMemory}.
Several stabilizer codes, including topological codes, have been experimentally implemented recently~\cite{Kelly2015StateCircuit, Takita2017ExperimentalQubits, Andersen2020RepeatedCode, Satzinger2021Realizing, Egan2021Fault-tolerantQubit, Chen2021ExponentialCorrection, Erhard2021EntanglingSurgery, Gong2021ExperimentalQubits, ryananderson2021realization, marques2021logicalqubit, Postler2021DemonstrationOperations, Krinner2021, Bluvstein2021}.

\begin{figure}
\centering
\includegraphics{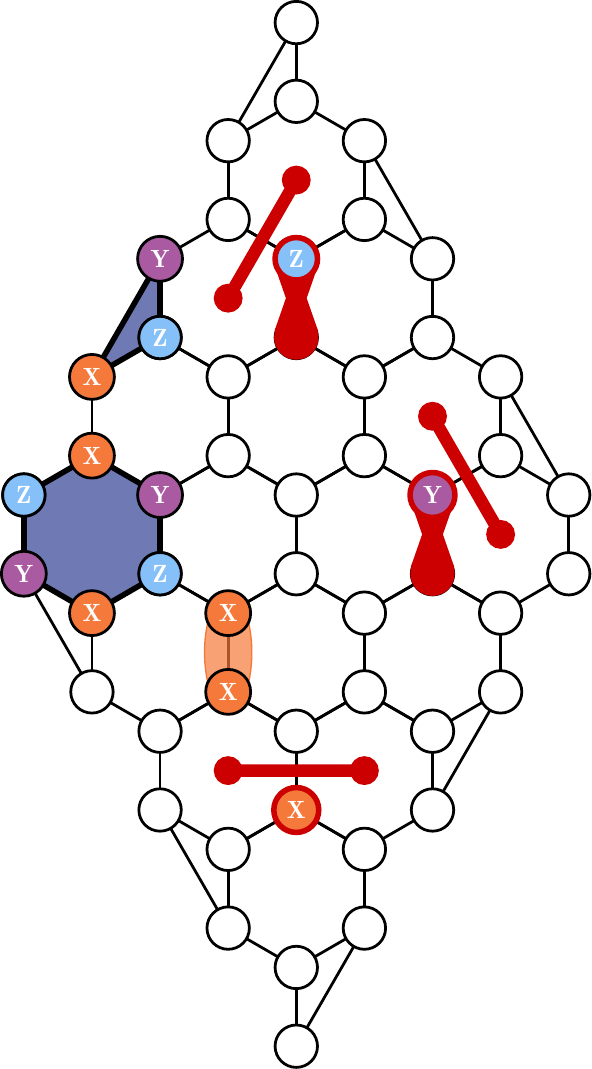}
\caption{The \phn code for distance $d = 5$. The stabilizers of the code are weight-six $XYZXYZ$ on each hexagonal plaquette, weight-two $XX$ on each vertical link, and weight-three $XYZ$ on the boundary. Also shown are the unidirectional syndromes (red) for isolated $X$, $Y$, and $Z$ errors.}
\label{fig:unidirectional syndromes}
\end{figure}

In this paper, we study a topological stabilizer code defined on a finite honeycomb grid. As shown in \figref{fig:unidirectional syndromes}, the code has weight-six stabilizers on the hexagonal plaquettes of the form $XYZXYZ$, and weight-two $XX$ stabilizers on ``vertical'' links. The code is the simplest example in the set of ``matching codes'' introduced by Wootton~\cite{Wootton2015}. For concreteness, we will refer to it as the \phn code, based on the structure of the plaquette and link stabilizers. We introduce boundaries with half-plaquette stabilizers, which yields a $[[2d^2,1,d]]$ code, i.e., it has code distance $d$ and one logical qubit for $2d^2$ data qubits. 

We study the basic properties of this code under the assumption of perfect stabilizer measurements and maximum-likelihood decoding. In particular, we compare the \phn code to the recently discussed XZZX code~\cite{Ataides2021XZZX}. The latter is a modification of the surface code~\cite{Bombin2007OptimalStudy, Tuckett2019Tailored}, where the original pure $X$ or pure $Z$ weight-four stabilizers are replaced by mixed $XZZX$ stabilizers on each plaquette of the square lattice. Similarly, the \phn code can be derived through a suitable set of single-qubit transformations on the rotated surface code, followed by a doubling of each qubit stabilized by weight-two operators. 

For depolarizing noise, up to the accuracy of our maximum-likelihood decoder, we find that the logical failure rate of these three codes depend in a similar way on the number of data qubits and the code distance. However, analogously to the XZZX code, but in contrast to the rotated surface code, the \phn code has very favourable properties for biased noise, with a threshold that increases with bias, reaching \unit[50]{\%} for pure noise. In fact, for phase-biased noise, the \phn code has lower sub-threshold logical failure rates, for the same number of data qubits, even though the two codes have the same threshold. The reason for the suppressed failure rate is that for pure $Z$ noise, the \phn code has code distance $2d^2$ while the XZZX code has code distance $d$.

Another interesting property of the \phn code is its distinct syndrome signatures (see \figref{fig:unidirectional syndromes}), which follow from the mixed plaquette stabilizers and the triangular structure. For isolated $X$, $Y$, and $Z$ errors, there are pairs of plaquette defects created, with unique orientations. We anticipate that this property may be used for approximate but efficient minimum-weight matching or clustering-type decoders. 

Although these results suggest that there are potential benefits of the \phn code compared to surface-type codes on a square grid, a potential disadvantage is that it requires high qubit connectivity to measure the weight-six stabilizers, which could make it more susceptible to measurement noise. On the other hand, this downside may be compensated by the fact that it also has weight-two stabilizers that may be possible to measure without ancilla qubits~\cite{Lalumiere2010, Kockum2012, Riste2013, Livingston2021TwoQubit, Gidney2021SimulationDGLT}. For similar number of data qubits, the \phn code has approximately half as many high-weight (plaquette) stabilizers as the XZZX code.

This article is structured as follows. In \secref{sec:XYZXYZ}, we first provide a more detailed account of the works and ideas leading up to the \phn code. Then we define the \phn code and some of its variations, and show how it can be constructed by a transformation of the rotated surface code. After explaining the methods used for error simulations and decoding in \secref{sec:Methods}, we present results in \secref{sec:Results}, comparing the logical failure rates and thresholds of the \phn and XZZX codes for different types of biased noise. We conclude and give an outlook for future work in \secref{sec:Conclusion}.


\section{The \texorpdfstring{\phn}{XYZ2} code}
\label{sec:XYZXYZ}

\subsection{Background}
\label{sec:Background}

The \phn code can be traced back to the Kitaev honeycomb model, which is an archetypical model for spin-liquid physics and topological order~\cite{Kitaev2006AnyonsBeyond, hermanns2018physics}. The model has spin-half degrees of freedom on every site of a honeycomb lattice with Ising-type interactions on links, but with a spin quantization axis that depends on the orientation of the link. The product of link operators around each hexagonal plaquette of the lattice commutes with the Hamiltonian, thus presenting a set of conserved quantities that can serve as stabilizers for an error-correcting code.

The most direct way to formulate an error-correcting code based on the Kitaev honeycomb model is to use the links as gauge operators of a subsystem code~\cite{Poulin2005StabilizerCorrection, Suchara2011ConstructionsCodes}. The centralizer of the gauge group, i.e., all group elements that commute with all elements in the group, make up the stabilizers of the code.  Measuring the link operators thus preserves the code space, corresponding to the stabilized subspace of the Hilbert space. As the stabilizers are part of the gauge group, these measurement outcomes can be constructed from link measurements. However, in order for the subsystem code to have any logical qubits, there has to be a set of logical operators that commute with the gauge group but that are not part of the gauge group. The link measurements must not change the logical state, but the measurements must also not measure the logical operators, since this would project the logical code word. Unfortunately, the subsystem code based on the link gauge operators can be shown to not have any logical qubits, so it is not directly useful as a quantum memory~\cite{Suchara2011ConstructionsCodes, Lee2017TopologicalModel}. 

Nevertheless, as shown recently by Hastings and Haah~\cite{Hastings2021DGLT, Haah2021BoundariesDGLT}, it is possible to dynamically generate logical qubits by a careful implementation of the cycle of measurements of link operators. This type of code has been found in simulations to be competitive with the surface code, especially for systems where the two-qubit link measurements do not require an additional ancilla qubit, thus reducing the detrimental effects of the large number of measurements~\cite{Gidney2021SimulationDGLT}.

Another alternative, presented by Wootton~\cite{Wootton2015, Wootton2017, Wootton2021}, is to formulate a stabilizer code on the hexagonal lattice by, in addition to the plaquettes, including as stabilizers strings of link operators between pairs of vertices, where each vertex is uniquely matched to a single other vertex. The simplest version of such ``matching codes'' is to match adjacent pairs of vertices along one direction, i.e., using one of the three link operators as a stabilizer. This is the \phn code, where our standard choice is to use the $XX$ links as stabilizers (but $YY$ or $ZZ$ links can be used instead, as discussed below).

As shown in Refs.~\cite{Wootton2015, Wootton2017}, introducing defects, by measuring a different link operator, logical qubits can be introduced and elegantly braided to perform logical operations. However, as discussed in the introduction, we will in this article only consider the properties of the code as a quantum memory, introducing boundaries and boundary stabilizers to construct a set of $[[2d^2,1,d]]$ codes. Furthermore, we only consider the ideal properties of these codes, assuming perfect stabilizer measurements. Even though this is not a realistic assumption for a real implementation of the code, using maximum-likelihood decoding and ideal measurements provides a baseline for the performance of the code. This baseline can be compared to that of other codes, independent of different schemes for modelling and decoding circuit-level noise.


\subsection{Construction of the code}
\label{sec:Construction}

As shown in \figref{fig:unidirectional syndromes}, the \phn code is constructed by placing $2d^2$ qubits on a hexagonal grid with a diamond-shaped boundary such that the setup contains $(d-1)^2$ plaquettes ($XYZXYZ$ stabilizers) and $d^2$ ``vertical'' links ($XX$ stabilizers). As we will see, the integer $d$ corresponds to the code distance, i.e., the minimum number of single-qubit operations required to perform a logical operation on a code word. We also introduce the notation $d_{\sigma}$, with $\sigma = X$, $Y$ or $Z$, to denote the distance of the code for a chain of only $X$, $Y$, or $Z$ operators. The qubits at the boundary are stabilized by an additional set of $2d-2$ weight-three half-plaquette $XYZ$ stabilizers, which are chosen such that no isolated single-qubit errors can go undetected. All together, these stabilizers make up the $2d^2 - 1$ constraints needed to encode a single logical qubit in $2d^2$ physical qubits. We note that the code can be truncated by removing qubits at the top and bottom of the grid, resulting in similar properties, but in this article, we only consider the full code as presented here. We also only consider odd $d\geq 3$, as the properties for even $d$ (as for the surface code) are different.

\begin{figure*}
\centering
\includegraphics[width=\linewidth]{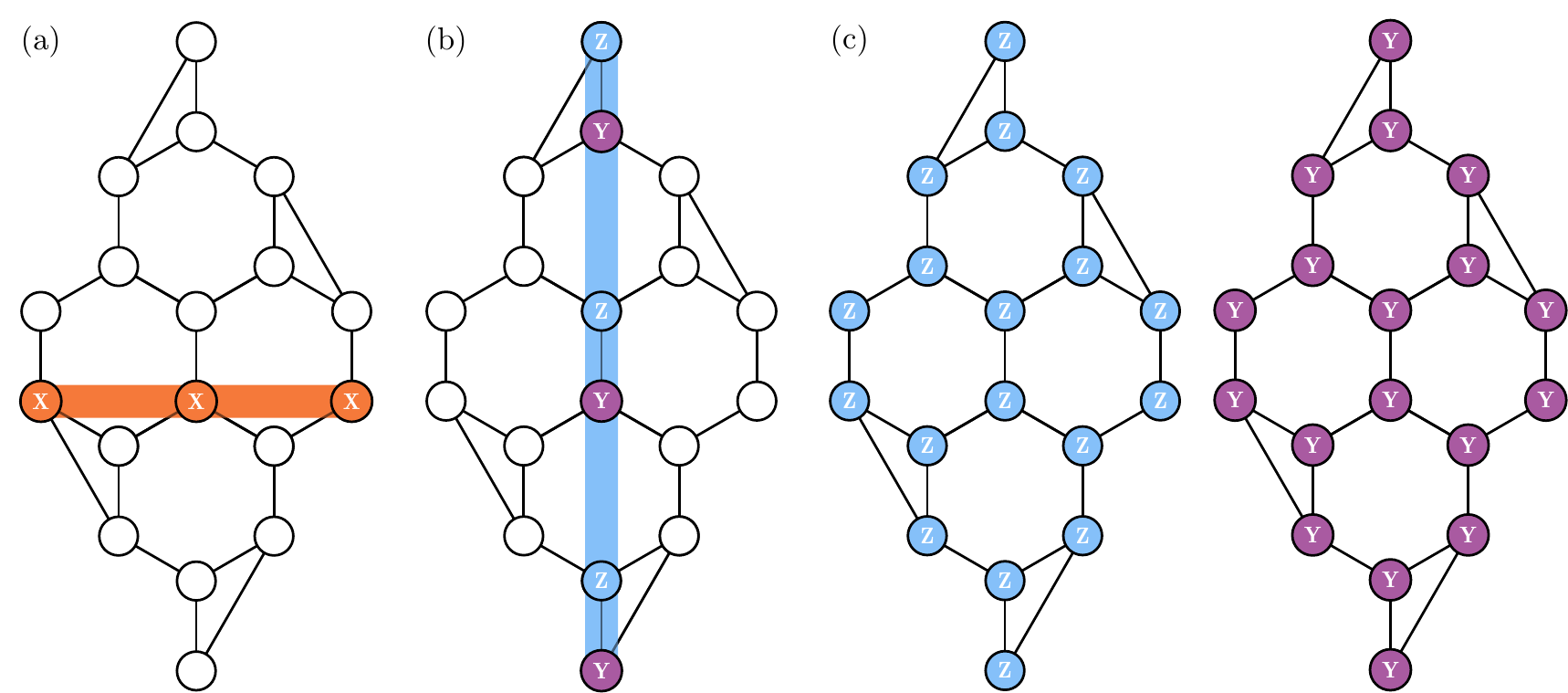}
\caption{Logical operators on a distance-3 \phn code.
(a) A pure logical $X$ operator $X_L$ formed by $X$s horizontally through the centre of the grid. This weight-$d$ operator is the shortest logical operator on the code and hence determines the code distance: $d_X = d$.
(b) A logical $Z$ operator $Z_L$ that goes vertically through the center of the grid. As this operator consists of both $Z$s and $Y$s, it is not a pure logical operator.
(c) Left: A chain of $Z$ operators applied on every qubit in the code constitutes a pure logical $Y$ operator $Y_L$. As this is the only pure operator consisting of $Z$s, it determines the code distance for pure $Z$ noise to be $d_Z = 2d^2$. Right: we can transform this operator by applying all $XX$ link stabilizers to obtain an operator consisting of $Y$s on all qubits. Thus we have $d_Z = d_Y = 2d^2$.}
\label{fig:logical operators}
\end{figure*}

The logical operators of a stabilizer code are the set of operators that commute with the stabilizers but are not part of the stabilizer group. These operators can thus operate on the code space without the corresponding changes being detectable by the stabilizer measurements. Any set of three anticommuting operators $X_L$, $Y_L$, and $Z_L$, defined up to deformations with stabilizers, provide the logical Pauli group.  In the \phn code, the shortest chain of single-qubit operators that commutes with all stabilizers is the horizontal chain of $d$ $X$s through the center of the grid [see \figpanel{fig:logical operators}{a}]. We designate this logical operator $X_L$ and infer from its existence that the code distance is $d$. As this chain of $X$s lies on $d$ qubits which are part of $XX$ vertical link stabilizers, the logical operator can be deformed by applying these $d$ stabilizers, giving a total of $2^d$ possible distance-$d$ pure $X_L$ operators. In addition, there is a large number of longer $X_L$ operators of mixed type, generated by acting with other stabilizers.

We define the logical $Z$ operator $Z_L$ as the vertical chain of $ZY/YZ$ link operators on the $2d$ qubits through the center of the grid, as shown in \figpanel{fig:logical operators}{b}. This chain can be deformed using adjacent stabilizers to form a mixed chain on the boundary of the code consisting of $X$, $Y$, and $Z$ errors. We note, however, that there is no representation of $Z_L$ which only has a single type of Pauli operator. Thus, for highly biased noise, this operator is unlikely to be generated through a random set of single-qubit errors. 

The logical $Y$ operator $Y_L$ can be expressed as a product of any representation of $Z_L$ and $X_L$. However, it turns out that we can deform this operator using plaquette and link stabilizers to create a pure-$Y$ logical operator that acts on all $2d^2$ qubits [\figpanel{fig:logical operators}{c}]. To see this, we note that the vertical links are made up of $XX$ stabilizers, which means that to construct a commuting operator made of only $Y$s, we need either a $YY$ or an $II$ ($I =$ Identity) operator on each vertical link. However, due to the plaquette and half-plaquette stabilizers consisting of $X$s, $Y$s, and $Z$s in a cyclic configuration, one needs a $YY$ on each link (or no link, but then it is trivial) in order for the logical operator to commute with each stabilizer. This argument works in a similar manner for a logical operator consisting of only $Z$ errors. In fact, these two logical operators are equivalent and can be transformed into each other by acting with all link stabilizers. In conclusion, a logical operator consisting of only $Z$ or $Y$ errors would need a chain of length $2d^2$ qubits. Hence the distance of the \phn code under pure $Z$ or $Y$ noise is $d_Z = d_Y = 2d^2$.

\begin{figure*}
\centering
\subfloat[]{
    \includegraphics[]{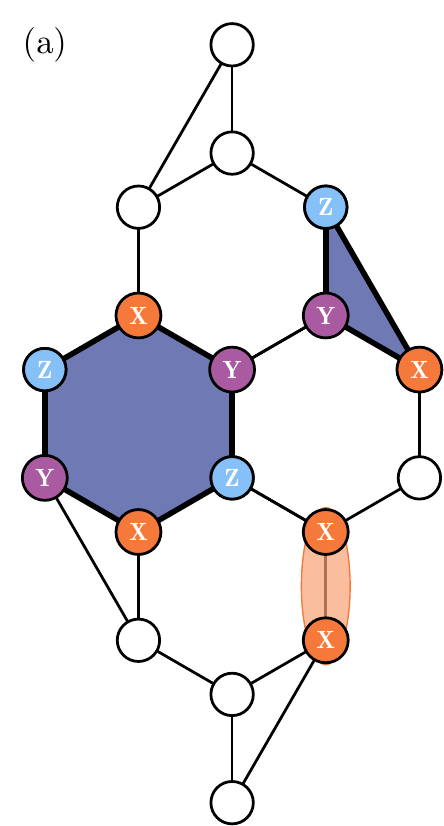}
    }
\hfill
\subfloat[]{
    \includegraphics[]{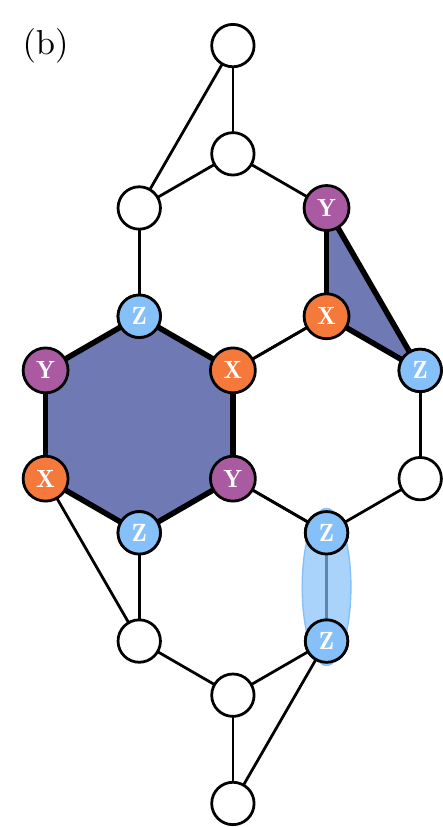}
}
\hfill
\subfloat[]{
    \includegraphics[]{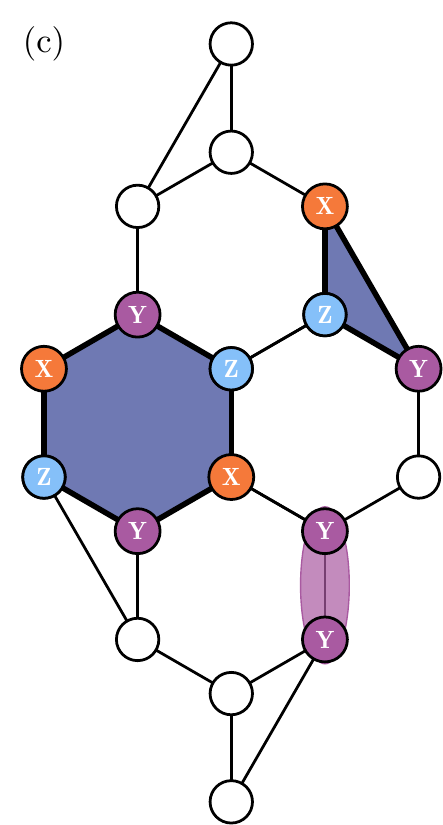}
}
\caption{Variations of the \phn code for distance 3 with (a) $XX$, (b) $ZZ$, and (c) $YY$ link stabilizers.}
\label{fig:variations of xyz code}
\end{figure*}

While we have chosen the vertical links to be $XX$ in this work, and hereafter assume this in all results shown unless stated otherwise, the code can be transformed into one using $ZZ$ or $YY$ link stabilizers by acting on the qubits with appropriate rotations. The logical operators will then transform accordingly as well. We show the different variations of the \phn code for distance 3 in \figref{fig:variations of xyz code}.

Similar to the XZZX code, the mixed nature and the uniformity of the plaquette stabilizers in the \phn code give rise to the desirable property that $X$ or $Z$ errors on single qubits result in syndromes in single, distinct directions. For the XZZX code, which lies on a square grid, these directions are the two diagonal directions. However, in the \phn code, this property is extended to cover $Y$ errors as well: $X$, $Y$, and $Z$ errors give rise to syndromes along their unique individual directions, as shown in \figref{fig:unidirectional syndromes}. For the XZZX code, this property was exploited by Ataides et al~\cite{Ataides2021XZZX} to build an efficient matching-based decoder. It would be interesting to explore whether the tri-directional nature of the syndromes for the \phn code can be similarly exploited to implement an efficient decoder. However, in contrast to the XZZX code and the surface code, where syndrome bits (except at the boundary) always appear in pairs, this is not the case for the link stabilizers of the XYZ$^2$ code, which means they are not directly amenable to minimum-weight matching. This is also in contrast to the honeycomb code, for which only the static stabilizer space, corresponding to plaquette stabilizers, is decoded \cite{Hastings2021DGLT,Gidney2021SimulationDGLT}.


\subsection{Transformation of the rotated surface code to the \texorpdfstring{XYZ$^2$}{XYZ2} code}
\label{sec:transformation_to_xyz}

As shown in Refs.~\cite{Kitaev2006AnyonsBeyond, kitaev2009topological,Vidal2008PerturbativeModel}  the low-energy sector of the honeycomb lattice model in the limit of large $ZZ$ coupling, replacing each such link by a single effective spin, is equivalent to the Wen model~\cite{Wen2003QuantumModel, Kay2011CapabilitiesMemory, Brown2011GeneratingMapping} with $YZZY$ interactions on a square lattice. This model can then be transformed to the toric code by applying individual rotations on the spins.
Here we perform similar transformations within the stabilizer formalism in reverse order, to show how the rotated surface code, the XZZX code, and the \phn code can be transformed into each other.

The rotated surface code is defined on a square lattice with alternating $XXXX$ and $ZZZZ$ plaquette stabilizers in the bulk and half-plaquette $XX$ and $ZZ$ stabilizers on the boundaries of the code. The shortest logical operators on this code are ``horizontal'' chains of $X$s that go across each column and ``vertical'' $Z$ chains that go across each row. In addition to these logical operators, there exist a pure-$X$ and a pure-$Z$ logical operator on each diagonal of the square grid. The only pure-$Y$ logical operator on the rotated surface code is the operator which acts on each of the $d^2$ qubits with a $Y$. This effectively makes the distance of the code $d$ under pure $X$ or $Z$ noise and $d^2$ under pure $Y$ noise.

The XZZX code can be obtained from the rotated surface code by applying a Hadamard rotation on every other qubit. This operation modifies both the $XXXX$ and $ZZZZ$ plaquette stabilizers of the rotated surface code to identical $XZZX$ stabilizers on each plaquette of the square lattice. As a consequence of the rotation, the shortest logical operators of the rotated surface code are transformed from pure $X$ and pure $Z$ chains to mixed chains, containing both $X$ and $Z$, going across horizontal and vertical dimensions of the grid. The two diagonal logical operators are exceptions, and still remain chains of pure $X$ or pure $Z$ errors. These last logical operators thus keep the code distance under pure $X$ or $Z$ noise $d$. The distance under pure $Y$ noise is $d^2$, as for the rotated surface code.

To transform to the \phn code, we start by doubling the number of qubits, with each pair stabilized by $XX$. The code space stabilized by this operator is spanned by $\ket{++}$ and $\ket{--}$, where $\ket{\pm} = \mleft( \ket{0} \pm \ket{1} \mright) / \sqrt{2}$. We make the choice of mapping the original single-qubit state $\ket{0}$ to $\ket{++}$ and $\ket{1}$ to $\ket{--}$. With this choice, the action of the Pauli operators on the single qubits is, in the two-qubit basis, mapped to
\begin{align}
    X &\xrightarrow{} ZZ \text{ or } YY , \label{eq1X} \\
    Y &\xrightarrow{} YZ \text{ or } ZY , \\
    Z &\xrightarrow{} XI \text{ or } IX .
\end{align}
%

\begin{figure}
\centering
\subfloat[]{
    \includegraphics[trim=0.25cm 0cm 0cm 0cm, clip=true]{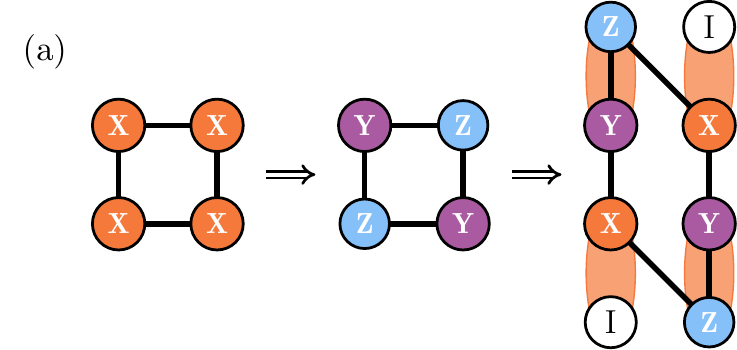}
    }
\\
\subfloat[]{
    \includegraphics[trim=0.25cm 0cm 0cm 0cm, clip=true]{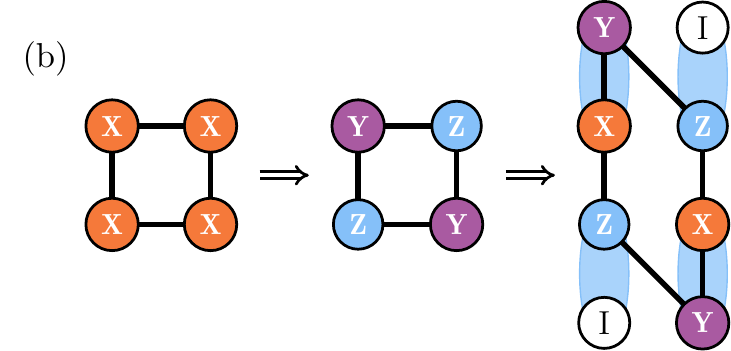}
}
\caption{Transforming $XXXX$ (and $ZZZZ$) plaquette stabilizers of the rotated surface code to variations of the \phn code.
(a) Transforming to the YZZY code by rotations and subsequently to the \phn code with vertical $XX$ link stabilizers by replacing each data qubit in the $YZZY$ plaquette with two qubits stabilized by $XX$.
(b) Transforming to the YZZY code by rotations and subsequently to the XYZ$^2$ code with vertical $ZZ$ link stabilizers by replacing each data qubit in the $YZZY$ plaquette with two qubits stabilized by $ZZ$.}
\label{fig:transformationXXandZZ}
\end{figure}

Now, starting from a slightly modified version of the XZZX code, with $YZZY$ plaquettes, this mapping results in the plaquette becoming a weight-six stabilizer ($XYZXYZ$), as shown in \figpanel{fig:transformationXXandZZ}{a}. Scaling this transformation up to the entire code gives us the \phn code [see \figpanel{fig:variations of xyz code}{a}].
The shortest horizontal and vertical logical operators on the YZZY code now become mixed chains containing $X$, $Y$, and $Z$, extending from one side to another. The diagonal weight-$d$ pure-$Z$ logical operator becomes a pure $X$ chain of length $d$, with $2^d$ degeneracy [\figpanel{fig:logical operators}{a}], whereas the orthogonal diagonal operator transforms to a $YZ$ logical operator of length $2d$ [\figpanel{fig:logical operators}{b}]. The pure-$X$ logical operator of the YZZY code acting on all qubits is mapped to pure $Z$ or $Y$ using \eqref{eq1X} [\figpanels{fig:logical operators}{c}{d}]. This ensures that while the distance of the \phn code under pure $X$ noise remains $d$, under pure $Y$ or $Z$ noise it becomes $2d^2$.

To obtain the variation of the \phn code shown in \figpanel{fig:variations of xyz code}{b}, where the vertical link stabilizers are $ZZ$ instead of $XX$, a similar transformation can be applied, as shown in \figpanel{fig:transformationXXandZZ}{b}. In this case, plaquettes from the rotated surface code are transformed into $YZZY$ plaquettes and then further into $XYZXYZ$ hexagonal plaquette stabilizers and $ZZ$ vertical link stabilizers by replacing each data qubit in the $YZZY$ plaquette by a pair of qubits stabilized by $ZZ$, such that $\ket{0}$ is represented by $\ket{00}$ and $\ket{1}$ is represented by $\ket{11}$.


\section{Methods}
\label{sec:Methods}

In order to characterize quantum error-correcting codes, we need to mimic, to an extent, the actual noisy conditions of an experiment. To do so, we simulate the code under various different noise models and observe its performance. In this work, we assume perfect stabilizer measurements and ancilla readouts, and focus only on single-qubit errors that may occur on the data qubits.

The error model we use is the standard asymmetric depolarizing channel,
\begin{equation}
    \rho \xrightarrow{} (1-p) \rho + p_x X\rho X + p_y Y\rho Y + p_z Z\rho Z \,,
\end{equation}
where $p = p_x + p_y + p_z$ is the total error probability per qubit per syndrome measurement cycle and $\rho$ is the density matrix describing the state of the code. This error channel introduces an $X$ error on a qubit with probability $p_x$, a $Y$ error with probability $p_y$, and a $Z$ error with probability $p_z$. If the probabilities of $X$, $Y$, and $Z$ errors occurring is equal, we call the error channel a symmetric depolarizing channel, and the noise simply depolarizing noise.

For phase-biased noise, we define
\begin{equation}
    \eta = \frac{p_z}{p_x + p_y} \hspace{1cm} (p_x = p_y)
\end{equation}
following Refs.~\cite{Tuckett2018UltrahighNoise, Tuckett2019Tailored, Ataides2021XZZX}. The limit $\eta \rightarrow \infty$ corresponds to pure $Z$ noise and $\eta = 0.5$ to depolarizing noise. We can correspondingly define bit-flip-biased ($X$-biased) or $Y$-biased noise. We note that $X$-biased noise for our standard version of the \phn code with $XX$ link stabilizers corresponds to $Z$-biased noise for the version with $ZZ$ link stabilizers.

In practice, for simulating the code, we do not need to consider the full density matrix, but only generate sets of single-qubit errors, so-called error chains, by sampling from the error distribution. From a single such chain, one chain in each of four equivalence classes of chains is generated by acting with each of the logical operators and the identity operator. These four chains are then deformed by acting with random stabilizers to preserve only the syndrome of the initial chain. A maximum-likelihood decoder~\cite{Dennis2002TopologicalMemory, Bravyi2014EfficientCode, Wootton2012HighCode, Hutter2014EfficientCode,Hammar2021ErrorRateAgnostic} (MLD) then calculates which of the four equivalence classes of chains is most likely to correspond to the syndrome. The event is counted as a successful error correction if the initial chain is in this most likely class, and a failed error correction otherwise. By sampling a large number of random initial chains, the logical failure rate of the code for a given error rate and noise bias is approximated.  

Any practical MLD is approximate, since a full count of all errors consistent with a random syndrome grows exponentially with the number of stabilizers and thus is infeasible to evaluate exactly for anything but very low error rates or very small codes.
In this work, we use the decoder recently presented in Ref.~\cite{Hammar2021ErrorRateAgnostic}, the ``effective weight and degeneracy'' (EWD) decoder, which is based on Monte Carlo sampling of error configurations using the Metropolis algorithm. 
This decoder is essentially a more efficient, simplified version of the Markov-chain Monte Carlo decoder formulated in Refs.~\cite{Hutter2014EfficientCode, Wootton2012HighCode}, which we have adapted to the \phn code and biased noise. As discussed in Ref.~\cite{Hammar2021ErrorRateAgnostic}, the EWD requires the tuning of a hyperparameter, corresponding to the error rate $p_{\text{sample}}$ (which acts as an effective temperature) of the Metropolis sampling, to minimize the logical failure rate for a given physical error rate $p$. There is thus some systematic uncertainty in the results, in the sense that the results provide upper bounds on the failure rate. For the XZZX code, we use the matrix-product-state (MPS) decoder as implemented in Ref.~\cite{qecsim}. As shown in \cite{Hammar2021ErrorRateAgnostic}, the two methods give almost identical results for the XZZX code for moderate code distances.


\section{Results}
\label{sec:Results}

Here we present results that compare the logical failure rates and the corresponding thresholds of the \phn and XZZX codes under the assumption of perfect stabilizer measurements and maximum-likelihood decoding, for biased noise. 
When comparing the results for these codes, it is important to keep in mind that while the rotated surface code and the XZZX code utilize $d^2$ data qubits and $d^2-1$ ancilla qubits placed in a square grid for a code of distance $d$, the \phn code uses twice the amount of data qubits, $2 d^2$, and the same amount of ancilla qubits, $d^2-1$, to achieve the same distance (although this requires a triangular lattice of qubits, where an ancilla qubits sits at the centre of each hexagonal plaquette in \figref{fig:unidirectional syndromes} and connects to all its nearest neighbours). Nevertheless, we will see that the \phn code matches the threshold values of the XZZX code, while having lower sub-threshold failure rates for phase-biased noise for similar numbers of data qubits.  

We begin by deriving exact expressions for logical failure rates for pure $X$, $Y$, and $Z$ noise, and compare those to the corresponding expressions for the XZZX code. This provides a good foundation for understanding the later observations using a finite bias. We also present results for depolarizing noise.


\subsection{Analytical expressions for pure X, Y, and Z noise}
\label{sec:pure_noise}


We consider the version of the \phn code with $XX$ link stabilizers, $N = 2 d^2$ qubits, and the code distance $d$ odd. Pure $Z$ noise and pure $Y$ noise are then equivalent, as discussed in \secref{sec:Construction}. There is only one pure-$Z$ operator that commutes with all the stabilizers: the weight-$N$ operator over all qubits [see \figpanel{fig:logical operators}{c}]. Thus, for a syndrome corresponding to any pure-$Z$ error chain, there is only one additional, complementary, chain that has the same syndrome. Maximum-likelihood decoding implies that for any syndrome, the most likely of the two will provide the correction chain. In other words, whenever an error chain with more than $N/2$ errors occurs, the error correction will fail. The marginal case, $N/2$ errors, fails with \unit[50]{\%} probability.

The logical failure rate, considering all possible error chains for pure $Z$ or $Y$ noise with error rate $p<0.5$, is thus
\begin{align}
    P_{f,Z}(p)=&\sum_{n=N/2}^{N}\binom{N}{n}p^n(1-p)^{N-n} \nonumber \\
    &-\frac{1}{2}\binom{N}{N/2}p^{N/2}(1-p)^{N/2} \,.
    \label{Pl_zpure}
\end{align}
The total probability of the chains for which the error correction will succeed is given by the complementary sum, which can be rewritten as $P_s(p)=P_f(1-p)$. Thus, given that $P_f(p)+P_s(p)=1$, we find, independently of the size of the code, that $P_f(p=0.5)=0.5$. This is the maximal failure rate for pure noise. Since it is independent of code distance, \unit[50]{\%} is also the threshold error rate. For low error rates,
\begin{equation}
P_{f,Z}\sim p^{N/2}=p^{d^2} \,,
\end{equation}
reflecting the code distance $2d^2$ for pure $Z$ or $Y$ noise.
Both the surface code and the XZZX code have these same properties for pure $Y$ noise (for $d$ odd), for which there is only a single logical operator, which consists of a $Y$ on all $N=d^2$ qubits. The expression for the failure rate is the same as in \eqref{Pl_zpure}, 
except that the last term is missing and the sum runs from $\lceil N/2\rceil$ to $N$.

For pure $X$ noise, the story is somewhat more complicated. As discussed in \secref{sec:Construction}, there are $2^d$ weight-$d$ pure-$X$ logical operators along the central stabilizer row of the code [see \figpanel{fig:logical operators}{a}], with the degeneracy corresponding to acting with the $XX$ link operators on this central row. To find out which chains will lead to logical failure, we first note that we can group chains by the parity of errors on each link: the probability of no error or two errors (even parity) on a link is $p_e $, and the probability of a single error (odd parity) is $p_o$, with
\begin{align}
    p_e &= (1-p)^2 + p^2, \nonumber\\
    p_o &= 2p(1-p) \,.
    \label{p_oddeven}
\end{align}
Acting with a logical operator changes the parity on each central link. Thus all of the chains which have the same parity of errors on each link are in the same equivalence class.

Given the syndrome, maximum-likelihood decoding corresponds to picking an arbitrary correction chain in the most likely equivalence class. For $p<0.5$, $p_o<p_e$, which means that for a given syndrome, the set of chains that are in the equivalence class with more than $d/2$ odd-parity errors on the central row will fail. Whatever the error configuration is outside the central row is irrelevant, since these will be equivalent for the two equivalence classes. The failure rate for pure $X$ noise is thus given by
\begin{equation}
    P_{f,X}(p)=\sum_{n=\lceil d/2 \rceil}^d\binom{d}{n}p_o^n p_e^{d-n} \,.
    \label{Pl_xpure}
\end{equation}
with $p_{o/e}$ given by \eqref{p_oddeven}.
Just as for pure $Z$ and $Y$ noise above, $P_f(0.5)=0.5$, meaning that $p=0.5$ is the threshold also for pure $X$ noise. For low error rates,
\begin{equation}
P_{f,X} \sim p^{\lceil d/2 \rceil} \,,
\end{equation}
as expected for code distance $d$. 

For the surface code, there is no simple analytical expression for either pure $X$ or $Z$ noise known. This lack of analytics is due to the fact that there is a large number of pure logical operators corresponding to the pure-$X$ and -$Z$ stabilizers. For the XZZX code, on the other hand, there is only a single pure logical operator of each type, of length $d$, such that the failure rate is given by \eqref{Pl_xpure}, with $p_o = p$ and $p_e = 1-p$, for both pure $X$ noise and pure $Z$ noise. 

\begin{figure}
\centering
\includegraphics[trim=0.4cm 0cm 1cm 0cm, clip=true, width=\linewidth]{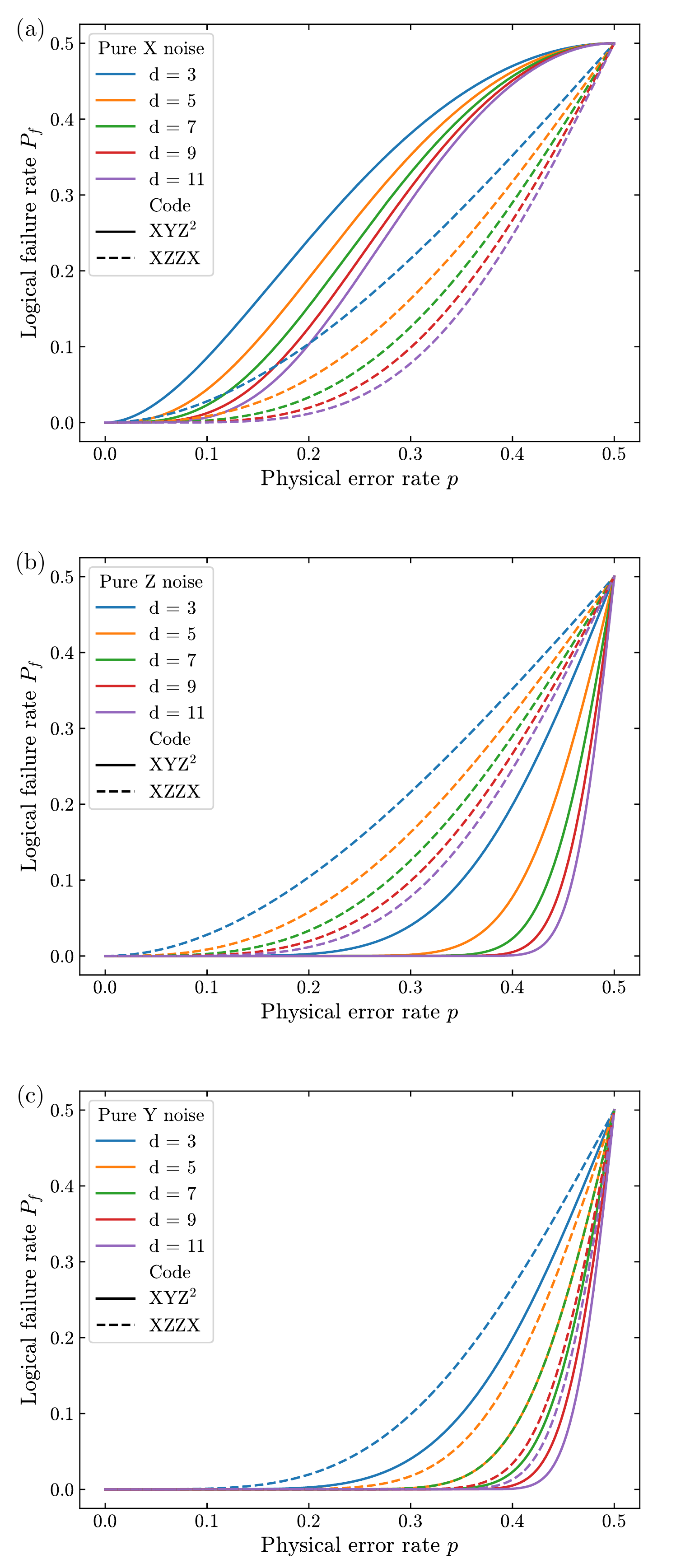}
\caption{Comparison of the logical failure rate $P_f$ for the \phn code (solid curves) and the XZZX code (dashed curves) as a function of the physical error rate $p$ for different code distances $d$ for
(a) pure $X$,
(b) pure $Z$, and
(c) pure $Y$ noise, plotted using analytical expressions, given in Eqs.~(\ref{Pl_zpure}) and (\ref{Pl_xpure}) for the \phn code.}
\label{fig:compareplot_purenoise}
\end{figure}

In \figref{fig:compareplot_purenoise}, we plot the pure-noise logical failure rates for a few low-$d$ code distances to visualize the differences between the \phn and XZZX codes. As shown above, both these codes have a \unit[50]{\%} error threshold for pure noise, where the logical failure rate $P_f(p=0.5)=0.5$ is the maximal error rate given that there are only two relevant equivalence classes. For pure bit-flip noise, even though the code distance is $d_X=d$ for both codes, the \phn code is clearly inferior to the XZZX code, which is a consequence of the $2^d$-fold degeneracy of the logical $X_L$ operator for the former. For pure phase-flip noise, the roles are reversed: here the $d_Z = 2d^2$ code distance of the \phn code yields logical fail rates that are significantly lower than those of the $d_Z = d$ XZZX code. Finally, for pure $Y$ noise, the two codes are equivalent (both to each other and to the rotated surface code), in the sense that the code distance is given by the total number of data qubits. Thus, e.g., the $d=5$ \phn code and the $d=7$ XZZX code overlap within the resolution of the plot. However, we note that, as discussed at the beginning of this section, for a given number of data qubits, the \phn code requires fewer ancilla qubits for stabilizer measurements than the XZZX code.


\subsection{Depolarizing noise}

\begin{figure}
\centering
\includegraphics[trim=0.7cm 0cm 1.5cm 1cm, clip=true, width=\linewidth]{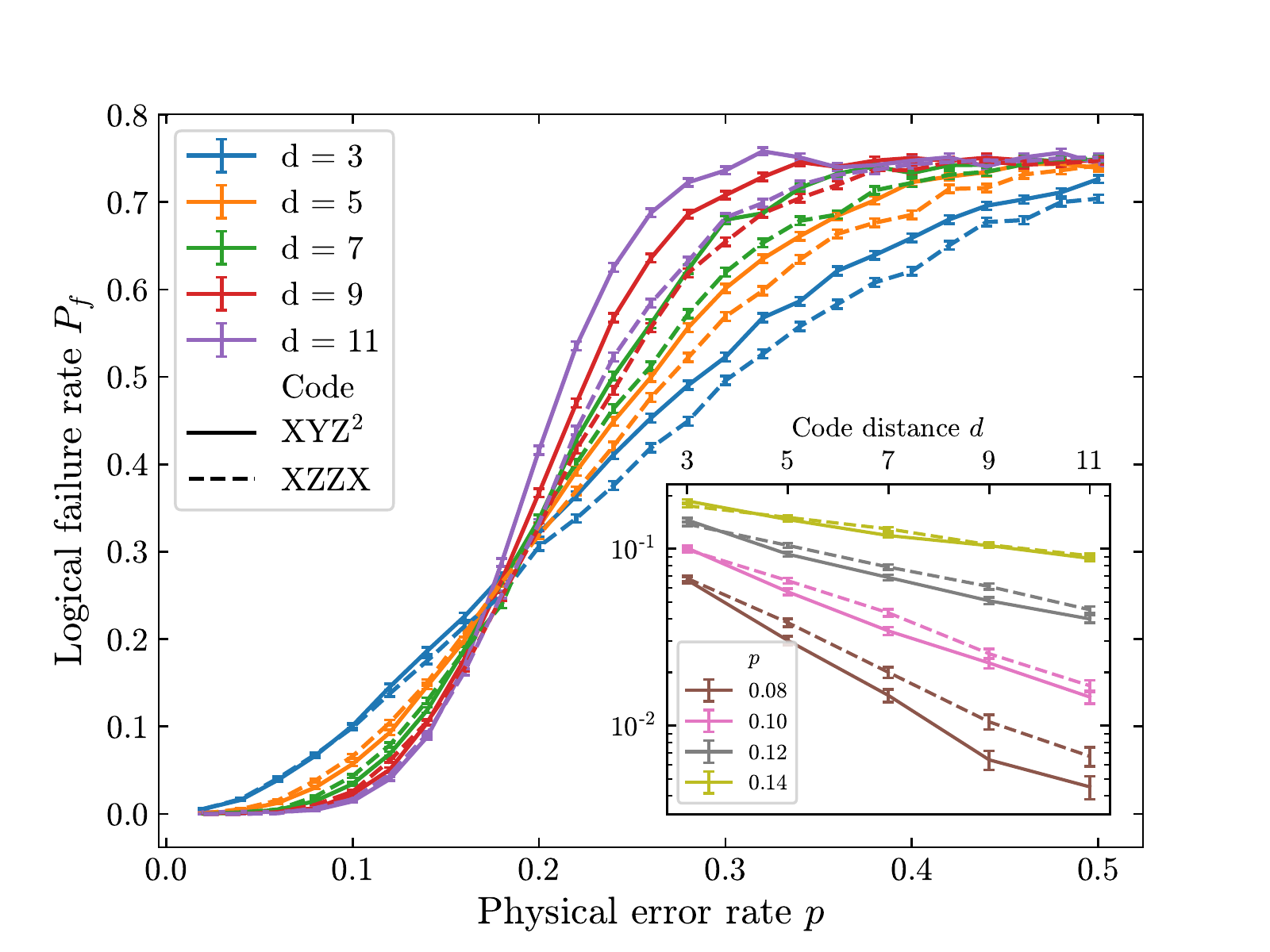}
\caption{Logical failure rates as a function of physical error rates for different code distances $d$ of the \phn (solid curves) and XZZX (dashed curves) codes for $\eta$ = 0.5 (depolarizing noise). Each data point is evaluated using 10,000 syndromes for each code, code distance, and physical error rate, using the EWD decoder for the \phn code and the MPS decoder for the XZZX code. The inset shows the logical failure rate as a function of code distance for a few error rates below the threshold $p\approx 0.18$. The solid and dashed curves are a guide to the eye, connecting subsequent data points. Error bars indicate the statistical error (one standard deviation) based on the number of sampled syndromes and the mean logical failure rate. We attribute larger-than-error-bar variations of the EWD decoder for large physical error rates to difficulties with optimizing the decoder's sampling error rate~\cite{Hammar2021ErrorRateAgnostic}.}
\label{fig:compareplot_dep}
\end{figure}

Before considering how the lessons from studying the model for pure noise translate to finite noise bias, we here compare the performance of the \phn and XZZX codes for depolarizing noise.
Logical failure rates for depolarizing noise on the two codes are shown as a function of the physical error rate in \figref{fig:compareplot_dep} for code distances between $d = 3$ and $d = 11$. The results for the \phn code are obtained using the EWD decoder (see \secref{sec:Methods}) with sampling error rate $p_{\text{sample}}$ increasing with $p$ and in the range $0.05$ to $0.6$, while the results for the XZZX code are obtained using the MPS decoder. 

Up to the accuracy of our decoders, we observe the same threshold, around \unit[18]{\%}, for the two codes. The logical failure rate is found to be a function of the number of data qubits above the threshold (for example, the lines for $d=5$ for the \phn code is here close to the line for $d=7$ for the XZZX, since this corresponds to almost the same number of data qubits) while for low error rates (inset of \figref{fig:compareplot_dep}) the logical failure rate instead scales with the code distance.


\subsection{Biased noise, \texorpdfstring{$\eta=10$}{eta=10}}

\begin{figure}
\centering
\includegraphics[trim=0.4cm 0cm 1cm 0cm, clip=true, width=\linewidth]{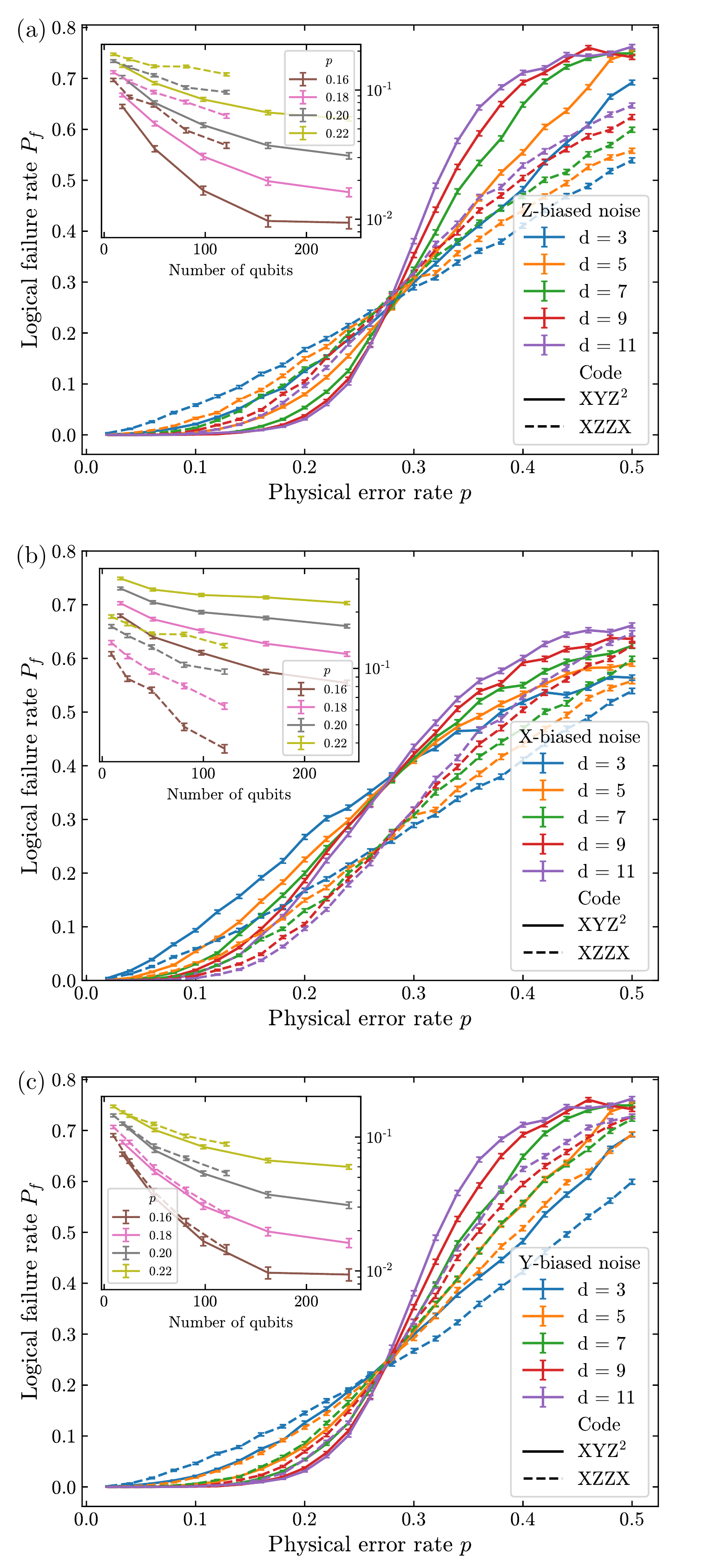}
    \caption{Logical failure rates as a function of physical error rates comparing equivalent code distances of the \phn code (solid curves) and the XZZX code (dashed curves) for $\eta$ = 10 for
    (a) $Z$-biased noise,
    (b) $X$-biased noise, and
    (c) $Y$-biased noise. The insets display the logical failure rates as a function of the number of data qubits for a few different sub-threshold error rates. For the \phn code, the same data is used in panels (a) and (c), since the code operates identically for $Z$- and $Y$-biased noise, and correspondingly for the XZZX code in panels (a) and (b). The number of syndromes used and the meaning of the error bars are the same as in \figref{fig:compareplot_dep}.}
    \label{fig:compareplot_eta10}
\end{figure}

We now explore finite-bias noise with $\eta = 10$, which corresponds to one dominant error channel, ten times more likely than the other two error channels combined. The logical failure rates for the \phn and XZZX codes under the three possible such noise biases are shown in \figref{fig:compareplot_eta10}. Results for the \phn code use the EWD decoder (with $p_{\text{sample}}\in [0.14,0.37]$ for $X$-biased noise and $p_{\text{sample}}\in [0.26,0.41]$ for $Z$- and $Y$-biased noise).

For $Z$-biased noise, the case shown in \figpanel{fig:compareplot_eta10}{a}, the threshold is about \unit[28]{\%} for both the \phn code and the XZZX code. However, we note that the logical failure rates below this threshold are highly suppressed for the \phn code compared to the XZZX code in this case. Comparing the results for an equivalent number of qubits instead of by code distance [inset in \figpanel{fig:compareplot_eta10}{a}], we observe still lower logical failure rates of the \phn code below the threshold.

In \figpanel{fig:compareplot_eta10}{b}, which shows the results for bit-flip-biased noise, the roles of the two codes are reversed: they exhibit similar thresholds, but the logical failure rates are significantly higher for the \phn code, consistent with a naive extrapolation from the pure-noise limit.

For $Y$-biased noise [\figpanel{fig:compareplot_eta10}{c}], the results for the \phn code are the same as for the case of $Z$-biased noise in \figpanel{fig:compareplot_eta10}{a}, while the logical failure rates for the XZZX code are suppressed below threshold compared to that case. As shown in the inset of \figpanel{fig:compareplot_eta10}{c}, the failure rate scales with the number of data qubits, again consistent with an extrapolation from pure noise. Thus, in summary, the results for $\eta=10$ show that the basic distinctions and similarities between the codes, based on the effective code distance for pure noise in \secref{sec:pure_noise}, survive also for moderate finite bias.


\section{Conclusion}
\label{sec:Conclusion}

We have studied a $[[2d^2,1,d]]$ stabilizer code, the ``XYZ$^2$'' code, defined on a honeycomb lattice with specific boundary conditions. The code can be derived as a concatenation of the surface code and a two-qubit phase-error-detection code, with suitable single-qubit rotations, and is the simplest realization of a ``matching code'' proposed by Wootton~\cite{Wootton2015}. The \phn code consists of weight-six $XYZXYZ$ stabilizers on hexagonal plaquettes, $XX$ link stabilizers on the vertical links, and $XYZ$ half-plaquette stabilizers on the boundary. Because of the hexagonal structure and the corresponding mixed-type stabilizers, the code has a remarkable syndrome signature, where isolated $X$, $Y$, and $Z$ errors have pairs of defects with three different respective orientations. 

We studied the logical fidelity of the \phn code assuming perfect stabilizer measurements and maximum-likelihood decoding, and compared the results to those of the rotated surface code and the XZZX code.
We found that the \phn code has high thresholds for biased-noise error models that far surpass those of the rotated surface code and matches the thresholds demonstrated by the XZZX code. In contrast to those two codes, the \phn code also has a quadratic code distance, $2d^2$, for both pure $Z$ and pure $Y$ noise. This distinction survives also for phase-biased noise with finite bias, where the \phn code has significantly lower sub-threshold logical failure rates for the same number of data qubits. We also noted that the \phn code requires fewer ancilla qubits per data qubits than the XZZX and rotated surface codes to carry out the stabilizer measurements, although this assumes higher connectivity: a triangular lattice of qubits instead of a square grid.

In order for these apparent advantages of the \phn code compared to the XZZX code to be of practical value, it will be necessary to device an efficient approximate decoder, potentially exploiting the tri-directional syndrome properties of the \phn code, while at the same time dealing with the fact that link stabilizers do not give rise to paired syndrome bits. A possible way to address the latter, which we are currently exploring, is to treat the \phn code as a concatenated code of two-qubit blocks~\cite{PhysRevA.74.052333}. A minimum-weight matching scheme could incorporate edges weighted by conditional probabilities based on the link syndrome~\cite{criger2018multi}. Something that also needs to be explored is the detrimental effects of the weight-six stabilizer on circuit-level noise~\cite{Wootton2021}, while taking into account that this may be partially offset by the fact that approximately half of the stabilizers are instead weight-two. In short, a complete analysis of the \phn code should be undertaken that relaxes the assumptions of perfect stabilizer measurements used in the present work. Furthermore, since the \phn code is especially apt at addressing biased noise, it would also be interesting to explore how the code would perform compared to the XZZX code as a concatenated code with Kerr-cat continuous-variable qubits~\cite{PRXQuantum.2.030345}.


\begin{acknowledgments}

We thank Karl Hammar for support using the EWD decoder and Timo Hillmann and Ben Criger for valuable discussions. We acknowledge the financial support from the Knut and Alice Wallenberg Foundation through the Wallenberg Centre for Quantum Technology (WACQT). Computations were enabled by resources provided by the Swedish National Infrastructure for Computing (SNIC) and Chalmers Centre for Computational Science and Engineering (C3SE). The software for running the EWD decoder for the \phn code is available on the repository~\cite{git}.

\end{acknowledgments}

\bibliography{ref}


\end{document}